\documentclass[11pt]{article}

\usepackage[preprint]{AIC}

\usepackage{times}
\usepackage{latexsym}

\usepackage{indentfirst} 
\usepackage{graphicx}
\usepackage{booktabs}
\usepackage[table,xcdraw,dvipsnames]{xcolor}
\usepackage{cuted}
\usepackage{caption}
\usepackage{hyperref} 
\usepackage{natbib}

\usepackage[T1]{fontenc}
\usepackage[utf8]{inputenc}
\usepackage{microtype}
\usepackage{inconsolata}

\usepackage{hyperref}
\usepackage{url}
\usepackage{authblk}

\title{The AI Committee: A Multi-Agent Framework for Automated Validation and Remediation of Web-Sourced Data}

\author[1,2,3]{Sunith Vallabhaneni}
\author[2,3]{Thomas Berkane}
\author[2,3]{Maimuna Majumder}

\affil[1]{UC Berkeley}
\affil[2]{Harvard Medical School}
\affil[3]{Boston Children's Hospital}

\affil[ ]{\texttt{\small sunithv@berkeley.edu, \{Thomas.Berkane, Maimuna.Majumder\}@childrens.harvard.edu}}

\date{} 

\begin{document}

\maketitle

\begin{abstract}
Many research areas rely on data from the web to gain insights and test their methods. However, collecting comprehensive research datasets often demands manually reviewing many web pages to identify and record relevant data points, which is labor-intensive and susceptible to error. While the emergence of large language models (LLM)-powered web agents has begun to automate parts of this process, they often struggle to ensure the validity of the data they collect. Indeed, these agents exhibit several recurring failure modes—including hallucinating or omitting values, misinterpreting page semantics, and failing to detect invalid information—which are subtle and difficult to detect and correct manually. To address this, we introduce the \textbf{AI Committee}, a novel model-agnostic multi-agent system that automates the process of validating and remediating web-sourced datasets. Each agent is specialized in a distinct task in the data quality assurance pipeline, from source scrutiny and fact-checking to data remediation and integrity validation. The AI Committee leverages various LLM capabilities—including in-context learning for dataset adaptation, chain-of-thought reasoning for complex semantic validation, and a self-correction loop for data remediation—all without task-specific training. We demonstrate the effectiveness of our system by applying it to three real-world datasets, showing that it generalizes across LLMs and significantly outperforms baseline approaches, achieving data completeness up to \texttt{78.7\%} and precision up to \texttt{100\%}. We additionally conduct an ablation study demonstrating the contribution of each agent to the Committee's performance. This work is released as an open-source tool for the research community.
\end{abstract}

\section{Introduction}

The development of robust models for core NLP tasks like event extraction and knowledge base population is fundamentally dependent on high-quality, structured data. While benchmark datasets have been invaluable, they are often static and cannot capture the dynamics of emergent, real-world events. This is a critical limitation for timely research in fields such as computational social science, public health, and economics, which rely on monitoring current affairs—from social unrest to disease outbreaks.

For such domains, the web is the essential, albeit noisy and unstructured, source of information. However, collecting comprehensive research datasets often demands manually reviewing many web pages to identify and record relevant data points, a process that is labor-intensive and susceptible to error. While the emergence of Large Language Model (LLM)-powered web agents has begun to automate parts of this process, they often struggle to ensure the validity of the data they collect. Existing agents are prone to subtle yet critical errors, ranging from the hallucination of plausible-sounding values to the misinterpretation of complex web page semantics. This forces researchers to validate each collected data point, undermining the initial promise of automation.

To address this, we introduce the \textbf{AI Committee} (AIC), a novel model-agnostic multi-agent system that automates the process of validating and remediating web-sourced data. This framework can be applied both to human-collected data and, most importantly, to data gathered by autonomous AI web agents, a rapidly growing source of data. Our system is designed as a digital assembly line, where each agent acts as a specialist with a narrowly defined role, mimicking a human-led data curation team.

Our framework integrates various in-context learning techniques, including few-shot learning for rapid dataset adaptation and structured chain-of-thought reasoning for complex semantic validation. For each dataset, context agents dynamically tailor AIC's validation rules before executing a comprehensive pipeline for each data point. This pipeline includes a source scrutinizer to assess source reliability, a fact-checker equipped with detailed, context-aware guidelines, and specialized agents to ensure data points are both relevant and structurally sound. A key feature of the system is its structured self-correction and discovery mechanism, where dedicated remediation agents correct invalid data points and discover new, relevant information from the source web pages. Additionally, AIC is model-agnostic and requires no model training, as it performs well across all models regardless of complexity—making it well-suited for low-resource academic/research settings.

To demonstrate the effectiveness of this framework, we test it on three datasets collected from the web by an LLM-based tool and use carefully human-cleaned versions as ground truth to evaluate accuracy, precision, and recall across several foundational models while analyzing the trade-off between performance, cost, and time. Our system significantly improves data quality across multiple state-of-the-art LLMs. This framework has the potential to accelerate the curation of high-quality datasets, empowering researchers to conduct more timely and impactful research.

\section{Related Work}

\subsection{Multi-Agent Systems for Web Data Collection}
Multi-Agent Systems (MAS) are rapidly emerging as the dominant paradigm for automating the complex, end-to-end task of web data collection. Recent state-of-the-art frameworks such as that of \citet{berkane2025llm} and AutoData \citep{ma2025auto} exemplify this trend. Both systems aim to transform a single natural language instruction into a final, structured dataset. AutoData employs a sophisticated architecture of 'research' and 'development' squads to programmatically generate and execute scraping code, featuring a novel hypergraph cache system to optimize inter-agent communication. Similarly, the work by \citet{berkane2025llm} details an end-to-end, human-in-the-loop pipeline that automates query generation, web page retrieval with bias mitigation, and data extraction. Other related systems like AutoScraper \citep{huang2024auto} focus on a similar problem domain, generating reusable wrappers (scrapers) for specific websites through a progressive, multi-stage process.

However, a critical analysis of these collection-focused systems reveals a common gap in the final, crucial stage: automated quality control and remediation. While they are powerful engines for initial data extraction, their approach to data quality largely remains a supervised or manual task. For instance, the framework by \citet{berkane2025llm} concludes its pipeline by using an LLM to flag potential data point issues for subsequent manual user review. While AutoData and
AutoScraper include validation steps, their focus is on the integrity and reusability of the
generated program or scraper, rather than the deep semantic validation and automated
correction of each individual extracted data point against its original source.

\subsection{LLMs for Data Quality and Cleaning}
There is a growing body of research on leveraging LLMs for data quality tasks such as data imputation, error detection, and schema mapping. For instance, \citet{zhang2024cocoon} explored the potential for LLMs to automate semantic data cleaning tasks—such as correcting inconsistent string representations—that were traditionally performed by complex regex rules or crowdsourcing. Moving beyond static cleaning, \citet{bendinelli2025exploring} investigated the use of LLM agents equipped with code execution tools (IPython) to iteratively detect errors and clean tabular datasets, specifically optimizing for downstream machine learning model performance.

While powerful, these approaches often focus on structural consistency or statistical optimization within a closed dataset. Our work differentiates itself by decomposing the complex problem of web-sourced data quality—which requires external validation against source text—into a series of distinct sub-tasks, each handled by an agent with a highly specialized prompt and purpose. This specialization, we argue, leads to more robust and interpretable outcomes than a single, monolithic approach.

\subsection{Prompt Engineering and In-Context Learning}
The performance of LLMs is heavily dependent on the quality of the prompt. Advanced techniques like chain-of-thought (CoT) prompting, which encourages the model to ``think step-by-step,'' have been shown to significantly improve performance on complex reasoning tasks \citep{wei2022chain}. Our \texttt{FactCheckerAgent} employs a highly structured prompt that not only enforces CoT reasoning but also incorporates a ``Critical Semantic Audit'' to actively check for logical fallacies. Furthermore, our framework's \texttt{ContextGenerator} embodies the principles of in-context learning, first popularized by \citet{brown2020language}. By analyzing a few data samples, these agents generate dynamic rules and tailored examples that are injected into the prompts of other agents, effectively conditioning the system's behavior for a specific dataset without updating model weights.

\subsection{Self-Correction and Automated Remediation}
A key frontier in LLM research is enabling models to reflect on and correct their own outputs. Many self-correction methods involve a simple feedback loop where the model critiques its own response. The AI Committee implements a more sophisticated, \textit{structured self-correction mechanism}. When the \texttt{ArbiterAgent} rejects a data point, the reason for rejection is passed to a dedicated \texttt{DataRemediationAgent}, a process inspired by self-refinement techniques described by \citet{madaan2023self}. This agent's sole purpose is to diagnose the failure and perform a targeted correction, creating a rigorous, multi-step validation loop that is more structured than a simple self-critique.

\section{The AI Committee}

\begin{figure*}[htb]
  \centering
  \includegraphics[width=0.95\textwidth]{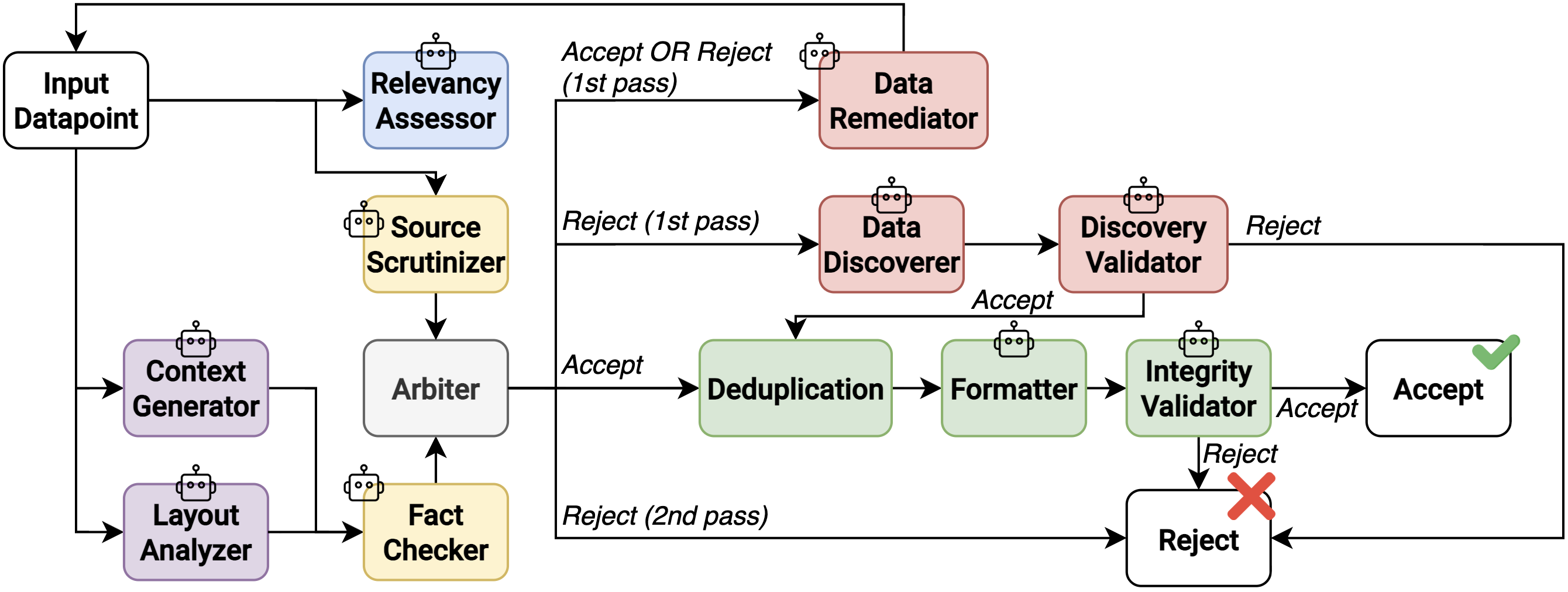}
  \caption{The data processing pipeline of AIC. Data points flow through a series of validation and remediation agents. Each box represents a step of the pipeline.}
  \label{fig:framework}
\end{figure*}

The AI Committee is a modular, asynchronous framework designed to clean either human or LLM web-sourced data points through a multi-stage validation and remediation pipeline. The overall architecture is depicted in Figure~\ref{fig:framework}, illustrating the flow of each data point through a series of specialized agents. To operate, the framework requires three user-provided inputs: 1) the initial tabular dataset of web-sourced data points with an associated source URL column for each data point, pointing to the web page from which the datapoint originates, 2) a high-level, natural language description of the dataset's purpose (e.g., "natural disaster events in Haiti and Cameroon"), and 3) a list of the schema — the dataset’s columns — to be validated (e.g., `event\_type`, `date`). These inputs form the basis for the Committee's context-aware processing.

\subsection{System Initialization and Adaptation}

Before processing any data, the framework undergoes an initialization phase to adapt itself to the specificities of the dataset it will operate on. This step employs few-shot learning principles: the framework provides the LLM with a small number of example datapoints directly within its prompt, instructing it to generalize from these examples to create a tailored operational context.

The following components are involved in this phase:

\textbf{ContextGenerator} This agent is responsible for creating the framework's operational context. It first analyzes a random subset of 10 rows of the input data to infer its semantic properties, mapping each schema field to: (1) a description of its expected entity type and granularity (e.g., \texttt{state}: 'geographic regions at the U.S. state level'); (2) a description of the temporal context (e.g., the date an event occurred); and (3) a list of negative examples to avoid (e.g., extracting a city for a \texttt{state} field). The number of negative examples is determined by the LLM's analysis of a field's complexity; a field with high ambiguity (e.g., a free-text \texttt{event\_type}) may warrant more examples than a well-defined one (e.g., \texttt{country}). Building on this analysis, the agent then generates richer, pedagogical illustrations of logical fallacies, providing the later \textbf{Fact Checker agent} with a deeper, context-specific understanding of not just what to avoid, but why.

\textbf{Dynamic Schema Generation.} From the list of user-provided schema fields, the framework programmatically generates a structured data format. For example, for a dataset tracking corporate acquisitions with fields \texttt{acquiring\_company}, \texttt{deal\_value}, and \texttt{date}, the system generates a schema requiring each data point to have an \texttt{acquiring\_company} (string), a \texttt{deal\_value} (integer), and a \texttt{date} (string, formatted as YYYY-MM-DD). This serves as a structure that all agents must adhere to when creating or modifying data. For instance, any new data point discovered by the \textbf{Data Remediation agent} must conform to this schema. In this context, structurally consistent is defined by the ability of a data point to be successfully parsed by this model, ensuring all required fields are present.

\subsection{Core Validation Pipeline}

Each data point from the input dataset is processed through the core validation pipeline, which functions as a digital assembly line.

\textbf{Relevancy Assessor Agent.} This agent performs an initial, low-cost check to determine if the data point is topically relevant to the dataset description. This check operates solely on the data point and the dataset description, occurring before any time-consuming web crawling or expensive content analysis is initiated. This prevents the system from wasting resources on fundamentally incorrect entries, as the data point will be rejected if it is found not to be relevant.

\textbf{Content Retrieval and Analysis.} For each relevant data point, the \textbf{LayoutAnalyzerAgent} uses Crawl4AI to parse the page layout into structured markdown. It then analyzes this markdown to classify the page's structure into one of several predefined categories, including \texttt{'ARTICLE'}, \texttt{'DIRECTORY\_LISTING'}, \texttt{'SEARCH\_RESULTS'}, \texttt{'HOMEPAGE'}, \texttt{'ERROR\_PAGE'}, or \texttt{'OTHER'}. The classification is used to provide a crucial analysis hint to the downstream Fact Checker agent.

\textbf{Source Scrutinizer Agent.} Concurrently with content retrieval, this agent analyzes the source URL to assess its trustworthiness, and then categorizes the source type and assigns a reliability score based on domain reputation.	

\textbf{Fact Checker Agent.} As the centerpiece of the validation pipeline, this agent is responsible for core semantic validation. Its prompt is dynamically assembled using the outputs from several upstream agents to provide rich context: it receives an analysis hint from the \texttt{LayoutAnalyzerAgent}, which provides a dynamically generated instruction in the agent's prompt to tailor its reading strategy. (eg: if a page is classified as a \texttt{'DIRECTORY\_LISTING'}, the hint explicitly warns the agent not to dismiss the page as purely navigational and to treat the text within list items as potential data points), a set of dataset-specific rules and tailored fallacy examples from the \texttt{ContextGenerator}. These dynamic inputs are combined with a prompt that includes a Critical Semantic Audit section that defines a set of logical principles hard-coded into the framework. This process involves extracting full entities instead of partial fragments, matching the data's granularity to the dataset's needs, analyzing qualifying terms, and identifying matches based on underlying meaning rather than just specific keywords. The agent's output is a structured `FactCheckerResponse` object, which contains boolean flags for content validity and factual accuracy, any newly extracted date, and detailed explanatory notes justifying its reasoning.

\textbf{Arbiter Agent.} Receives the structured reports from the \texttt{FactCheckerAgent} and the \texttt{SourceScrutinizerAgent}. It functions not as an LLM call, but as a deterministic, rule-based decision engine. Evidence is collated through a series of logical checks: a data point is immediately rejected if the \texttt{FactCheckerAgent} reports that the source text lacks meaningful content or does not support the data point's claims. It is also rejected if the \texttt{SourceScrutinizerAgent} deems the source's reliability to be 'Low' or 'Very Low'. Only if a data point passes all of these checks does the Arbiter issue a verdict of \texttt{'ACCEPT'}.

\textbf{Data Formatter Agent.} Upon receiving an \texttt{'ACCEPT'} verdict, this agent performs a final structural normalization pass. It ensures the data point strictly adheres to the generated schema types and formats (e.g., coercing stringified numbers to integers, enforcing ISO date standards) before the data is passed to the finalization stage.

\subsection{Remediation and Discovery: A Structured Self-Correction Loop}

If the \texttt{ArbiterAgent} rejects a datapoint, it is not immediately discarded. Instead, it enters the remediation loop, which enables correction and data discovery.

\textbf{Data Remediation Agent.} This agent attempts to fix the rejected data point. It is capable of two types of fixes: direct value replacement (e.g., correcting an incorrect city name that is explicitly mentioned in the text) and calculation-based remediation. The process begins with an "Analyst" LLM call, which receives the rejected data point, the reason for rejection, and the source text, and in turn produces a structured remediation plan. If this plan requires external information for a calculation (e.g., finding a state's total population to apply a percentage found in the text), it can dispatch a fact-lookup tool. This tool queries Google, scrapes the top results, and uses an LLM to extract the required information. If the remediation is successful, the newly corrected datapoint is validated by a Remediation Auditor—a modified Fact Checker specialized in verifying correction logic—before being finalized. If it fails, the datapoint is permanently rejected.

\textbf{Data Discovery Agent.} This agent scans the entire source page to discover if other, different data points matching the schema are present. This allows the framework to augment the dataset with new relevant information from the same sources.

\subsection{Finalization}

Datapoints that are accepted, either initially or after remediation, pass through two final stages:

\textbf{Hierarchical Deduplication.} A deterministic algorithm performs a final cleaning of the accepted datapoints. The process begins by filtering out any record with missing values in the required schema fields. If a `date` field is present, the algorithm creates a "base fingerprint" for each datapoint from all non-date fields. Based on that fingerprint, it then deduplicates entries by checking for this fingerprint at three distinct levels of date granularity: full date (YYYY-MM-DD), year-month, and year-only. The levels can coexist because a record is only checked against the set corresponding to its own precision; for instance, accepting a record with month-level precision does not preclude accepting a different record for the same event with day-level precision. This allows otherwise identical records with differing date specificities to coexist. If no `date` field is in the schema, a standard deduplication across all fields is performed instead, which identifies duplicates based on identical values across all user-defined schema fields and retains only the first unique instance encountered.

\textbf{Data Integrity Validation Agent.} This agent performs a last-pass quality check on the deduplicated data. It is explicitly prompted to validate each data point against only two simple, hard-coded rules: (1) \textbf{Completeness:} Is any required schema field empty, null, or missing? While the preceding \textbf{Hierarchical Deduplication} step also filters for missing fields, it does so merely as a structural prerequisite to enable its algorithm. This agent's check, in contrast, serves as a final quality control on the deduplicated output. (2) \textbf{Plausibility:} Does any field contain a value that is obvious nonsense given its description (e.g., a numerical value for a field requiring a person's name)? Data points that fail this final check are discarded.

\section{Demonstration Description}
\label{sec:demonstration}

We demonstrate AIC via an interactive web interface built with Streamlit, designed to provide a transparent view into the autonomous validation process. The demonstration highlights the system's end-to-end functionality, from raw data ingestion to a cleaned, validated output.

The user workflow goes through the following steps:

\textbf{1. Configuration:} The user uploads a CSV file containing web-sourced data points, provides a natural language description of the dataset, specifies the schema fields or columns to be validated, and inserts their OpenAI API key.

\textbf{2. Execution:} Upon starting the process, the system begins processing each data point through the multi-agent pipeline.

\begin{figure*}[t]
  \centering
  \includegraphics[width=\textwidth]{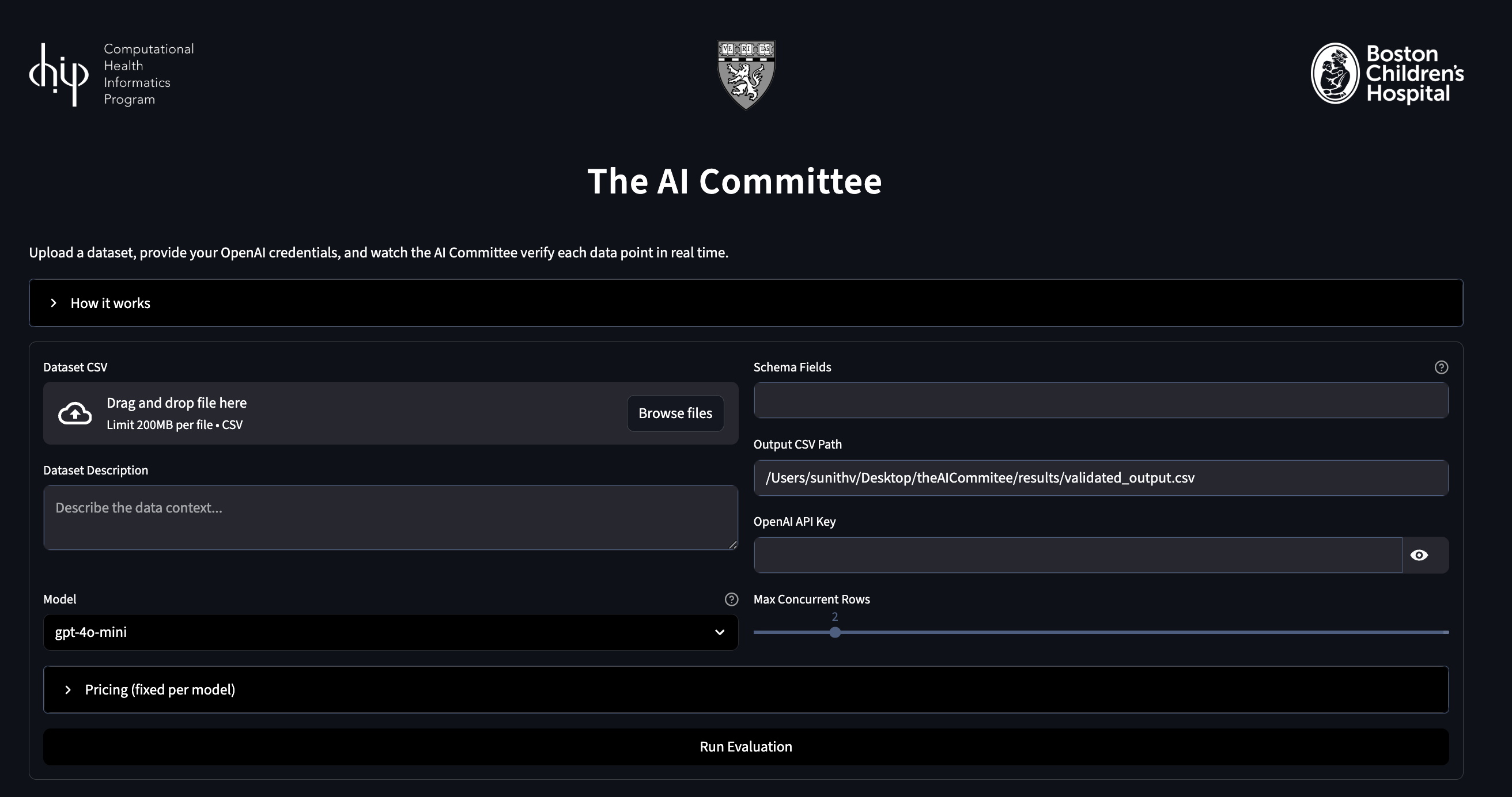}
  \caption{The web interface of AIC. Users provide a dataset, description, and schema, then monitor the real-time validation status of each data point as the system runs.}
  \label{fig:web}
\end{figure*}

\textbf{3. Real-Time Monitoring:} The interface displays a real-time status log, as shown in Figure~\ref{fig:web}. Each data point is listed with its current status (e.g., \texttt{Processing}, \texttt{ACCEPT}, \texttt{REJECT}, \texttt{DISCOVERED}), allowing the user to observe the Committee's decisions as they happen.

\textbf{4. Results:} Once the pipeline completes, the final, validated dataset is made available for download, and a summary of the performance and cost metrics is displayed.

\section{Evaluation, Availability \& Licensing}
\label{sec:evaluation}

We conduct a comprehensive empirical evaluation to validate the effectiveness and analyze the operational characteristics of the AIC. Our experiments, orchestrated by a reproducible evaluation harness, are designed to measure performance across multiple dimensions, from data quality and computational cost to the granular performance of each agent in the pipeline.

\subsection{Datasets}

To establish a high-quality ground truth, each dataset was independently annotated by three public health researchers following a detailed inter-annotator agreement (IAA) protocol, which will be released alongside our code.

\begin{enumerate}
    \item \textbf{Natural Disasters:} 125 initial data points (i.e., LLM-collected) related to natural disasters in Haiti and Cameroon.
    \item \textbf{Police Misconduct:} 95 initial data points on police misconduct incidents in the United States.
    \item \textbf{COVID-19 Tracing:} 83 initial data points on the number of downloads of COVID-19 contact tracing applications per state in the US.
\end{enumerate}

\subsection{Systems Under Test}
To rigorously evaluate the framework, we compare several architectures and model configurations:

\begin{itemize}
    \item \textbf{The AI Committee (AIC):} The modular, multi-agent framework as described in Section 3. We evaluate the Committee using three different underlying models to test generalizability and cost-efficiency: \texttt{gpt-4o-mini}, \texttt{o4-mini}, and \texttt{gpt-5}.
    
    \item \textbf{Monolithic Agent Baseline:} A single agent powered by the same models, provided with the full schema, dataset description, and a complex, all-in-one prompt. This baseline serves to demonstrate the specific value of decomposing the task into specialized agents versus a standard "zero-shot" or "few-shot" LLM approach.
    
    \item \textbf{Rule-Based Baseline:} A deterministic system that applies a series of hard-coded heuristics, regex patterns, and data cleaning rules specific to each dataset. This represents traditional, non-LLM data cleaning methods.
    
    \item \textbf{Ablation Configurations:} We systematically disable key agents (e.g., Relevancy Assessor, Fact Checker, Remediation) within the AI Committee to quantify the individual contribution of each component to the overall system performance.
\end{itemize}

\subsection{Results}

Our results demonstrate that the AIC framework significantly optimizes the trade-off between data quality, computational cost, and latency. Most notably, the AIC configuration powered by \texttt{gpt-4o-mini} achieved the highest overall performance (\textbf{F1: 85.1}), surpassing both larger, more expensive models \texttt{gpt-5} and \texttt{o4-mini}.

\begin{figure}[t!]
  \centering
  \includegraphics[width=\columnwidth]{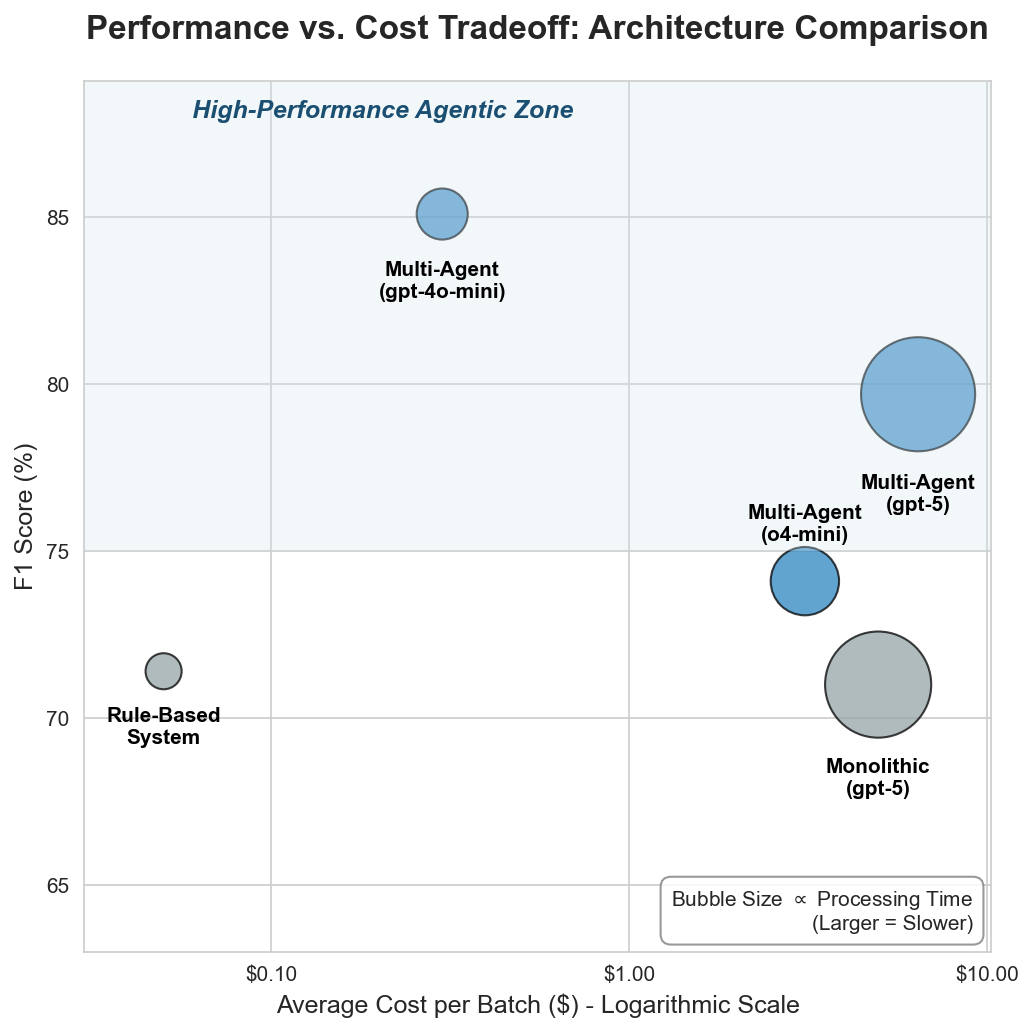}
  \caption{Performance vs Cost Tradeoff. Bubble size represents total processing time.}
  \label{fig:tradeoff}
\end{figure}

Figure~\ref{fig:tradeoff} visualizes this landscape. \texttt{gpt-4o-mini} dominates the ``high-performance, low-cost'' (top-left) quadrant, delivering a \textbf{14.1 point F1 improvement} over the Monolithic baseline while reducing operational costs by over 16$\times$ (\$0.30 vs \$4.95). Conversely, while the \texttt{gpt-5}-powered Committee achieved high precision (100.0\%), it had significantly higher cost and latency.

Table~\ref{tab:main_results} details these metrics. The modular architecture consistently outperforms the Monolithic baseline, which struggled with recall (58.9\%), likely due to attention dispersion over long contexts. The Rule-based baseline, while precise, failed to generalize to complex unstructured text, resulting in the lowest recall (58.2\%) of any valid configuration. The ablation study further validates our design: removing key components—such as the Remediation Agent or Fact Checker—resulted in measurable drops in F1 score (-6.4 and -1.3, respectively), confirming that high data quality requires a collaborative, multi-stage system rather than a single pass.

\begin{table}[h!]
    \centering
    \caption{\textbf{Architecture Evaluation}: F1, P=Precision, R=Recall, Rem=Remediation Recall. $\Delta$ values show percentage point change (Absolute Difference) from the AIC (4o-mini) baseline. Operational metrics (Time, Latency, Cost) are averaged per dataset batch.}
    \label{tab:main_results}

    \newcommand{\dpos}[1]{\textbf{\textcolor{ForestGreen}{+#1}}}
    \newcommand{\dneg}[1]{\textcolor{BrickRed}{#1}}
    \newcommand{\dneu}{\textcolor{gray}{--}}

    \small 
    \setlength{\tabcolsep}{2.5pt} 
    \renewcommand{\arraystretch}{1.2} 

    \rowcolors{3}{gray!10}{white}

    \resizebox{\columnwidth}{!}{%
        \begin{tabular}{l c c c c | c c c}
            \toprule
            & \multicolumn{4}{c|}{\textbf{Performance ($\Delta$)}} & \multicolumn{3}{c}{\textbf{Operations (Avg)}} \\
            \textbf{System} & \textbf{F1} & \textbf{P} & \textbf{R} & \textbf{Rem} & \textbf{Time} & \textbf{Lat} & \textbf{Cost}\\
            \midrule
            
            \rowcolor{blue!5} 
            \textbf{AIC (4o-mini)} &
            \textbf{85.1} (\dneu) &
            \textbf{92.7} (\dneu) &
            \textbf{78.7} (\dneu) &
            \textbf{77.8} (\dneu) &
            1,285s & 11.7s & \$0.30 \\

            AIC (o4-mini) &
            74.1 (\dneg{-11.0}) &
            92.7 (\dneu) &
            61.7 (\dneg{-17.0}) &
            61.1 (\dneg{-16.7}) &
            3,128s & 27.0s & \$3.09 \\

            AIC (gpt-5) &
            79.7 (\dneg{-5.4}) &
            100.0 (\dpos{6.3}) &
            66.7 (\dneg{-12.0}) &
            44.4 (\dneg{-33.4}) &
            9,258s & 88.9s & \$6.40 \\

            Monolith (gpt-5) &
            71.0 (\dneg{-14.1}) &
            89.5 (\dneg{-3.2}) &
            58.9 (\dneg{-19.8}) &
            58.3 (\dneg{-19.5}) &
            7,632s & 75.9s & \$4.95 \\

            Rule-Based &
            71.4 (\dneg{-13.7}) &
            92.4 (\dneg{-0.3}) &
            58.2 (\dneg{-20.5}) &
            55.6 (\dneg{-22.2}) &
            -- & -- & -- \\

            \bottomrule
        \end{tabular}%
    }
\end{table}

\begin{table}[h!]
    \centering
    \caption{\textbf{Ablation Study} Metrics: P=Precision, R=Recall, F1=F1 Score, Rem=Remediation Recall. $\Delta$ values show percentage point change (Absolute Difference) from the AIC (4o-mini) baseline.}
    \label{tab:ablation_singlecol}

    \newcommand{\dpos}[1]{\textbf{\textcolor{ForestGreen}{+#1}}}
    \newcommand{\dneg}[1]{\textcolor{BrickRed}{#1}}
    \newcommand{\dneu}{\textcolor{gray}{0.0}}

    \small 
    \setlength{\tabcolsep}{2.5pt} 
    \renewcommand{\arraystretch}{1.2} 

    \rowcolors{3}{gray!10}{white}

    \resizebox{\columnwidth}{!}{%
        \begin{tabular}{l c c c c}
            \toprule
            \textbf{System} & \textbf{P ($\Delta$)} & \textbf{R ($\Delta$)} & \textbf{F1 ($\Delta$)} & \textbf{Rem ($\Delta$)}\\
            \midrule
            
            \rowcolor{blue!5} 
            \textbf{AIC (4o-mini)} &
            \textbf{92.7} (\dneu) &
            \textbf{78.7} (\dneu) &
            \textbf{85.1} (\dneu) &
            \textbf{77.8} (\dneu) \\

            No FactCheck &
            92.4 (\dneg{-0.3}) &
            76.6 (\dneg{-2.1}) &
            83.8 (\dneg{-1.3}) &
            58.3 (\dneg{-19.5}) \\

            No Context &
            93.0 (\dpos{0.3}) &
            74.5 (\dneg{-4.2}) &
            82.7 (\dneg{-2.4}) &
            75.0 (\dneg{-2.8}) \\

            No CtxExamples &
            95.3 (\dpos{2.6}) &
            71.6 (\dneg{-7.1}) &
            81.8 (\dneg{-3.3}) &
            63.9 (\dneg{-13.9}) \\

            Rem-Only &
            95.3 (\dpos{2.6}) &
            70.9 (\dneg{-7.8}) &
            81.3 (\dneg{-3.8}) &
            66.7 (\dneg{-11.1}) \\
            
            No Integrity &
            90.5 (\dneg{-2.2}) &
            73.8 (\dneg{-4.9}) &
            81.3 (\dneg{-3.8}) &
            66.7 (\dneg{-11.1}) \\

            No SrcScrutiny &
            91.4 (\dneg{-1.3}) &
            73.0 (\dneg{-5.7}) &
            81.2 (\dneg{-3.9}) &
            66.7 (\dneg{-11.1}) \\

            No CtxLearning &
            92.8 (\dpos{0.1}) &
            71.6 (\dneg{-7.1}) &
            80.8 (\dneg{-4.3}) &
            66.7 (\dneg{-11.1}) \\

            No Remediation &
            95.9 (\dpos{3.2}) &
            66.7 (\dneg{-12.0}) &
            78.7 (\dneg{-6.4}) &
            58.3 (\dneg{-19.5}) \\

            No Layout &
            94.1 (\dpos{1.4}) &
            67.4 (\dneg{-11.3}) &
            78.5 (\dneg{-6.6}) &
            61.1 (\dneg{-16.7}) \\
            
            Discovery-Only &
            96.0 (\dpos{3.3}) &
            66.0 (\dneg{-12.7}) &
            78.2 (\dneg{-6.9}) &
            58.3 (\dneg{-19.5}) \\
            
            Min FactCheck &
            83.6 (\dneg{-9.1}) &
            65.2 (\dneg{-13.5}) &
            73.3 (\dneg{-11.8}) &
            58.3 (\dneg{-19.5}) \\

            No Relevancy &
            78.9 (\dneg{-13.8}) &
            68.1 (\dneg{-10.6}) &
            73.1 (\dneg{-12.0}) &
            66.7 (\dneg{-11.1}) \\

            No Formatter &
            92.1 (\dneg{-0.6}) &
            41.1 (\dneg{-37.6}) &
            56.9 (\dneg{-28.2}) &
            50.0 (\dneg{-27.8}) \\
            \bottomrule
        \end{tabular}%
    }
\end{table}
\subsection{Limitations and Future Work}

Our framework, while powerful, has several limitations. \textbf{Dependence on LLM Capabilities:} The ultimate performance is bounded by the reasoning and knowledge capabilities of the underlying LLM. Hallucinations or misinterpretations by the LLM can still lead to errors. \textbf{Prompt Optimization Bias:} We acknowledge that the prompts employed across our agents were iteratively developed and refined using \texttt{gpt-4o-mini} as the primary testbed. Due to this, the superior cost-performance ratio observed for this specific model may  result from prompt-model alignment, where instructions are inadvertently optimized for its specific reasoning patterns. Other models, such as \texttt{gpt-5} or \texttt{o4-mini}, likely achieve higher performance ceilings if the prompts were specifically re-tuned for their respective attention mechanisms and instruction-following quirks. \textbf{The Messiness of Web Data:} The framework is still susceptible to the inherent ``messiness'' of the web. Paywalls, complex JavaScript-rendered pages, and highly unconventional metadata can impede content retrieval and analysis. \textbf{Ground Truth Subjectivity:} Our evaluation is based on a human-curated ground truth which, particularly for date fields, involved a degree of subjective leniency due to the ambiguous nature of dates in web contexts. A different ground truth definition could alter the perceived accuracy of the models. \textbf{Reliance on LLM's Internal Knowledge:} A potential limitation of our framework is the \texttt{SourceScrutinizerAgent}'s reliance on the LLM's internal knowledge to assess source reliability, which carries an inherent risk of hallucination. We argue, however, that this risk is minimal for this specific task. The agent's function---classifying major domains like news outlets or government portals---relies on stable, high-consensus knowledge that is deeply encoded in the training data of any large foundation model. Furthermore, our empirical results support this design choice: in our experiments across all three datasets and models, we observed no instances where the agent hallucinated or mischaracterized a source's type or reputation.

Future work will focus on two key areas. First, we plan to conduct a more thorough benchmark of available models from different families (e.g. Claude, Gemini, etc.) and explore hybrid approaches, using different models for different agents to optimize the cost-performance trade-off (e.g., a fast, cheap model for relevancy screening and a powerful, expensive model for fact-checking). Second, we will continue to refine the agents' reasoning capabilities through prompt engineering, particularly in handling ambiguity and complex temporal expressions, fixing edge cases as we find them in our testing and user feedback.

\subsection{Availability \& Licensing}
The AI Committee is released as an open-source project under the \texttt{Apache 2.0} license. The complete codebase, including our comprehensive evaluation, the three annotated ground truth datasets, and the inter-annotator agreement protocol, is available at:

\noindent
\textbf{Code Repository:} \url{https://github.com/sunith-v/theAICommitteeDemo}

\noindent
\textbf{Video:} \url{https://youtu.be/c4xI9F1s24E}

\bibliography{custom}

@inproceedings{berkane2025llm,
  title={LLM-based web data collection for research dataset creation},
  author={Berkane, Thomas and Charpignon, Marie-Laure and Majumder, Maimuna S.},
  booktitle={Findings of the Association for Computational Linguistics: EMNLP 2025},
  pages={12610--12622},
  year={2025},
  publisher={Association for Computational Linguistics}
}

@article{ma2025auto,
  title={AutoData: A multi-agent system for open web data collection},
  author={Ma, Tianyi and Qian, Yiyue and Zhang, Zheyuan and Wang, Zehong and others},
  journal={arXiv preprint arXiv:2505.15859},
  year={2025}
}

@inproceedings{huang2024auto,
  title={AutoScraper: A progressive understanding web agent for web scraper generation},
  author={Huang, Wenhao and Gu, Zhouhong and Peng, Chenghao and Liang, Jiaqing and others},
  booktitle={Proceedings of the 2024 Conference on Empirical Methods in Natural Language Processing},
  pages={2371--2389},
  year={2024}
}

@inproceedings{wei2022chain,
  title={Chain-of-Thought Prompting Elicits Reasoning in Large Language Models},
  author={Wei, Jason and Wang, Xuezhi and Schuurmans, Dale and Bosma, Maarten and others},
  booktitle={Advances in Neural Information Processing Systems},
  volume={35},
  pages={24824--24837},
  year={2022}
}

@inproceedings{brown2020language,
  title={Language Models are Few-Shot Learners},
  author={Brown, Tom and Mann, Benjamin and Ryder, Nick and Subbiah, Melanie and others},
  booktitle={Advances in Neural Information Processing Systems},
  volume={33},
  pages={1877--1901},
  year={2020}
}

@inproceedings{madaan2023self,
  title={Self-Refine: Iterative Refinement with Self-Feedback},
  author={Madaan, Aman and Tandon, Niket and Gupta, Prakhar and Hallinan, Skyler and others},
  booktitle={Advances in Neural Information Processing Systems},
  volume={36},
  year={2023}
}

@article{zhang2024cocoon,
  title={Data Cleaning Using Large Language Models},
  author={Zhang, Shuo and Huang, Zezhou and Wu, Eugene},
  journal={arXiv preprint arXiv:2410.15547},
  year={2024}
}

@article{bendinelli2025exploring,
  title={Exploring LLM Agents for Cleaning Tabular Machine Learning Datasets},
  author={Bendinelli, Tommaso and Dox, Artur and Holz, Christian},
  journal={arXiv preprint arXiv:2503.06664},
  year={2025},
  note={Proceedings of the ICLR 2025 Workshop on Foundation Models in the Wild}
}

\end{document}